\documentclass[
superscriptaddress,
preprint,
amsmath,amssymb,
 prl
]{revtex4-2}
\usepackage{graphicx}
\usepackage{dcolumn}
\usepackage{bm}
\usepackage{float}
\usepackage[utf8]{inputenc}

\begin{document}


\title{Electrons with Anomalous Energy Generated in Gas-Filled and Vacuum Diodes}

\author{Victor~F.~Tarasenko}
\author{Vasily~Yu.~Kozhevnikov}
\email[Corresponding author: ]{Vasily.Y.Kozhevnikov@ieee.org}
\author{Andrey~V.~Kozyrev}
\author{Evgenii~Kh.~Baksht}
\author{Maxim~S.~Vorobyov}
\affiliation{Institute of High Current Electronics SB RAS, Tomsk 634055, Russia}


\begin{abstract}
It is shown that, when high-voltage pulses (with a voltage amplitude exceeding 100~kV in centimeter gaps) with a leading edge duration of 1~ns or shorter are applied to gas-filled and vacuum electric discharge diodes, electrons with kinetic energies nominally exceeding the amplitude of the applied voltage are detected. In the experiments, electron beam attenuation curves were measured in absorbers consisting of Al foil of varying thicknesses. These curves were used to reconstruct the electron beam energy spectrum by regularizing the solution of an integral equation based on deep machine learning. The obtained spectra contain electrons with anomalously high energies, the proportion of which, depending on the conditions, can reach 25~\%. A control experiment with a long voltage pulse on a large-area vacuum diode (voltage 150 kV, pulse duration $35\ \mu$s, vacuum gap 12~cm, electrode area $75 \times 15\ cm^2$) showed that the proportion of electrons with anomalous energies is less than 0.2~\%. Experiments have shown that the main mechanism for generating electrons with anomalous energy is the spatio-temporal synchronism of the motion of fast electrons in the enhanced field formed in the gap by space charge.
\end{abstract}

\maketitle


\section{\label{sec:level1}Introduction}

It has now been reliably established that when high-voltage pulses are applied to vacuum \cite{Khudyakova, Bugaev, Shpak, Ganter} and gas-filled \cite{Tarasova, Babich, Baksht, Kozyrev_2015} diodes, electrons with energy $T$ exceeding $eU$ are sometimes detected behind the anode foil of the accelerators, where $e$ is the electron charge and $U$ is the amplitude of the voltage pulse across the gap. Theoretical studies \cite{Boichenko, Kozhevnikov_2022} show that, for vacuum diodes with nanosecond voltage pulses, the generation of high-energy electrons with $T > eU$ is due to the formation of a transient electron cloud, whose electric field further increases the electrons’ energy as they reach the anode. To explain the generation of anomalous-energy electrons (AEE) in gas-filled diodes, the “polarization” mechanism proposed by Askarian \cite{Askaryan1, Askaryan2} is commonly used; it involves accelerating electrons at the front of a fast ionization wave (streamer or spark leader). This idea was used by Babich \cite{Babich} to explain the appearance of AEE in air at atmospheric pressure during the formation of runaway electron (RE) beams. Runaway electrons are those that, while moving between collisions with gas atoms or molecules, gain more energy in the electric field than they lose in collisions. This mechanism was proposed by Wilson while studying atmospheric discharges \cite{Wilson} based on Rutherford’s formula \cite{Thomson}.

In air at atmospheric pressure, RE (including AEE) were first detected in \cite{Tarasova} by measuring the current through a resistive shunt. The same study also detected RE in other gases at various pressures. The possibility of generating AEE in air at atmospheric pressure was later confirmed in \cite{Alekseev}. In \cite{Kostyrya}, it was shown that in air at atmospheric pressure, the RE beam current can reach $100$~A due to a significant increase in the cathode's active area. RE beam currents in the tens of amperes were also obtained with tubular cathodes by applying a longitudinal magnetic field \cite{Mesyats}. The possibility of generating RE with anomalous energies was confirmed in \cite{Mesyats_2020, Huang_2025}. However, the fraction of of AEE in RE beam in gas-filled diodes was relatively small; for example, in \cite{Baksht, Kozyrev_2015}, it did not exceed $10$~\%.

In 2023, Pasko et al. \cite{Pasko_2023} suggested that during atmospheric discharges, AEE can be formed via Photoelectron Emission during the Absorption of Bremsstrahlung (PEAB). This mechanism was presented in detail in \cite{Pasko_2024}, where it was compared with experimental results from gas-filled diodes reported in \cite{Tarasova, Tarasenko_2020}. During PEAB in \cite{Pasko_2024}, X-ray bremsstrahlung quanta, arising from the bombardment of a metal anode by fast electrons, knocked out additional photoelectrons from the cathode. Moreover, some of these photoelectrons could have energies close to $eU$. Furthermore, these fast electrons (from subsequent generations) continue to gain kinetic energy in the discharge gap. In this way, their energy can exceed $eU$ by a factor of two or more over several cycles.

However, a preliminary analysis of our results \cite{Baksht, Kozyrev_2015, Kozhevnikov_2022, Zhang_2013, Tarasenko_2020} and data from other sources \cite{Tarasova, Babich, Boichenko, Mesyats_2020, Mesyats, Huang_2025} on RE generation of electron beams in gas-filled and vacuum diodes indicates that the physical mechanism proposed in \cite{Pasko_2023, Pasko_2024} cannot be the primary mechanism for gas-filled diodes \cite{Tarasova, Tarasenko_2020}. We also note that the term "runaway electrons," proposed in \cite{Dreicer_1959, Dreicer_1960} for processes in gases and plasmas, is not appropriate for describing electron motion in a vacuum.

This paper compares experiments with gas-filled and vacuum diodes that detect electrons with energies $T\ \geq\ eU$, using new methods for fast-electron-spectrum diagnostics \cite{Kozhevnikov_2026}. Reconstructing electron energy spectra from measured attenuation curves in foils reduces to solving an ill-posed inverse problem for the first-kind Fredholm integral equation. Classical regularization methods used previously \cite{Baksht, Kozyrev_2015} were insufficiently robust to errors in the kernel of the Fredholm equation (the Tabata-Ito formula \cite{Tabata_1975}). This paper presents a new technique for solving the complete ill-posed problem using deep machine learning. This technique accounts for uncertainties in both the attenuation curve and the operator kernel and allows processing of raw experimental data sets without additional interpolation or selection \cite{Kozhevnikov_2026}. This toolkit is the primary means of processing the experimental data in this paper.

\section{\label{sec:level2}Experimental setups and procedures}

Experimental measurements of electron-beam generation in gas-filled and vacuum diodes, and of their attenuation in foils, were conducted using three setups. Setup~1a utilized the SLEP-150M accelerator \cite{Tarasenko_2011}, to which an additional transmission line was connected, and a conical current collector was used to detect the electron beam behind the foil. The use of an additional transmission line made it possible, based on the incident and reflected voltage waves – as in \cite{Baksht, Kozyrev_2015, Kostyrya, Tarasenko_2020} – to reconstruct the voltage pulse over the interval and synchronize it with the beam current and the discharge current through the diode. In experiments with a gas-filled diode, a cathode in the form of a stainless steel sphere with a diameter of $9.5$~mm was used. The interelectrode gap was $6.5$~mm and filled with air at atmospheric pressure.

Measurements at Setup~1b were also conducted on the same Setup~1a, but with a vacuum diode. In this case, a cathode in the form of a cylinder with a diameter of $6$~cm, rolled from $50\ \mu$m-thick stainless steel foil, was used. The interelectrode gap was $4$~mm. Due to the absence of electron energy losses, the beam current in the vacuum was increased by an order of magnitude compared to that of a gas-filled diode. 

The SLEP-150M accelerator generator produced voltage pulses with a falling-edge amplitude of $130$~kV and a half-height duration of $1$~ns. In “idle” mode, the voltage amplitude across the diode doubled. The rise time of the voltage pulse was $250$~ps at the $0.1–0.9$ levels of amplitude. In both Setup~1, the beam current waveform was recorded. The objective was to determine the relative number of AEE and to measure the amplitude of the voltage pulse during beam current generation. To calculate the electron energy distribution, the corresponding attenuation curve was recorded. The electron beam was detected behind an anode filter made of aluminum foil of various thicknesses, reinforced with a diaphragm or mesh. Electron spectra were then calculated from the recorded attenuation curve using a new method for solving the ill-posed inverse problem. 

The beam current amplitude (or the number of electrons in the beam) was measured using three current collectors of different designs: i) a conical collector with a $20$-mm receiving-section diameter and a temporal resolution of $80$~ps; ii) a diaphragm with a $0.5$-mm diameter aperture was placed immediately behind the anode foil, followed by a collector with a $3$-mm receiving-section diameter and a temporal resolution of $25$~ps \cite{Kostyrya}; iii) a disc-shaped collector with a $56$-mm receiving-section diameter (used to measure the electron current across the entire foil surface); its resolution did not allow the temporal profile of a RE pulse with a duration of less than $1$~ns to be resolved. The total current in the diode during gas discharge and the short-circuit current were recorded using a shunt made of chip resistors \cite{Kozyrev_2015}. 

To determine the timing of AEE appearance in a vacuum diode, Setup~2 was used. The measurement circuit included a RADAN-220 generator that produced negative-polarity voltage pulses with a $2$~ns FWHM across a load consisting of an IMA3-150E vacuum tube. The electron beam current was measured using a low-resistance shunt made of chip resistors with a total resistance of $0.1\ \Omega$. The receiving part of the shunt consisted of either a $38$-mm-diameter disk electrode or a $20$-mm-diameter current collector connected to a cable with a characteristic impedance of $50\ \Omega$. The temporal resolution of the first sensor was no worse than $0.2$~ns, and that of the second was $80$~ps. The purpose of the experiments on Setup~2 was to determine the time of AEE generation relative to the current-pulse profile in the vacuum diode.

The voltage, beam current, and discharge current pulses in the gas-filled diode were fed to real-time digital oscilloscopes. During operation on Setups 1 \& 2, Tektronix TDS6604 oscilloscopes ($6$~GHz and $20$ samples/ns) and LeCroy WaveMaster 830Zi-A oscilloscopes with a bandwidth of up to $30\ GHz$ and a sampling rate of $12.5$~ps ($80$ samples/ns) were used. To record the beam current pulses, a 1-meter-long RG58-A/U (Radiolab) high-frequency cable was used, along with N-type (Suhner 11 N-50-3-28/133 NE) and SMA (Radiall R125.075.000) connectors. 

Setup~3 was based on the “DUET” \cite{Vorobyov_2015} long-pulse ($35\ \mu$~s) electron accelerator with a plasma cathode and a wide-aperture beam outlet. The plasma cathode’s flat section ($75 \times 15\ cm^2$) was covered with a fine-mesh screen grid ($0.4$~mm square) with $44$~\% light transmittance. Under a constant accelerating voltage applied between the plasma emitter grid and the foil exit window (a $12\ cm$ gap that serves as the anode of a high-voltage diode gap), electrons are extracted from the plasma and accelerated to energies of $eU$. The electron beam was extracted into the atmosphere through an exit window ($75 \times 15\ cm^2$) covered with a $40-\mu$~m-thick aluminum-beryllium composite foil. The foil was mounted on a support structure with a geometric transparency of $56$~\%. The voltage across the gap was measured with a high-impedance voltage divider, and the total current through the diode was measured with a Rogowski coil installed at the positive terminal of the capacitor bank. In the experiment, the voltage drop across the accelerating gap did not exceed $1$~\% at $150$~kV. A TDS-3034 oscilloscope ($0.3$~GHz, $2.5$ samples/ns) was used to record the current and voltage waveforms in Setup~3. A shielded current collector with a $56$~mm diameter receiving section has been placed $10$~mm from the anode foil to capture electrons emitted into the atmosphere. 

The thickness of the attenuating filter was adjusted by stacking multiple layers of aluminum foil. As the total filter thickness increased, the collector current decreased monotonically and eventually stabilized, after which the signal became indistinguishable from the electrical noise level.

\section{\label{sec:level2}Results of beam current measurements and electron spectra calculations}

\begin{figure}[h]
\includegraphics[scale=0.32]{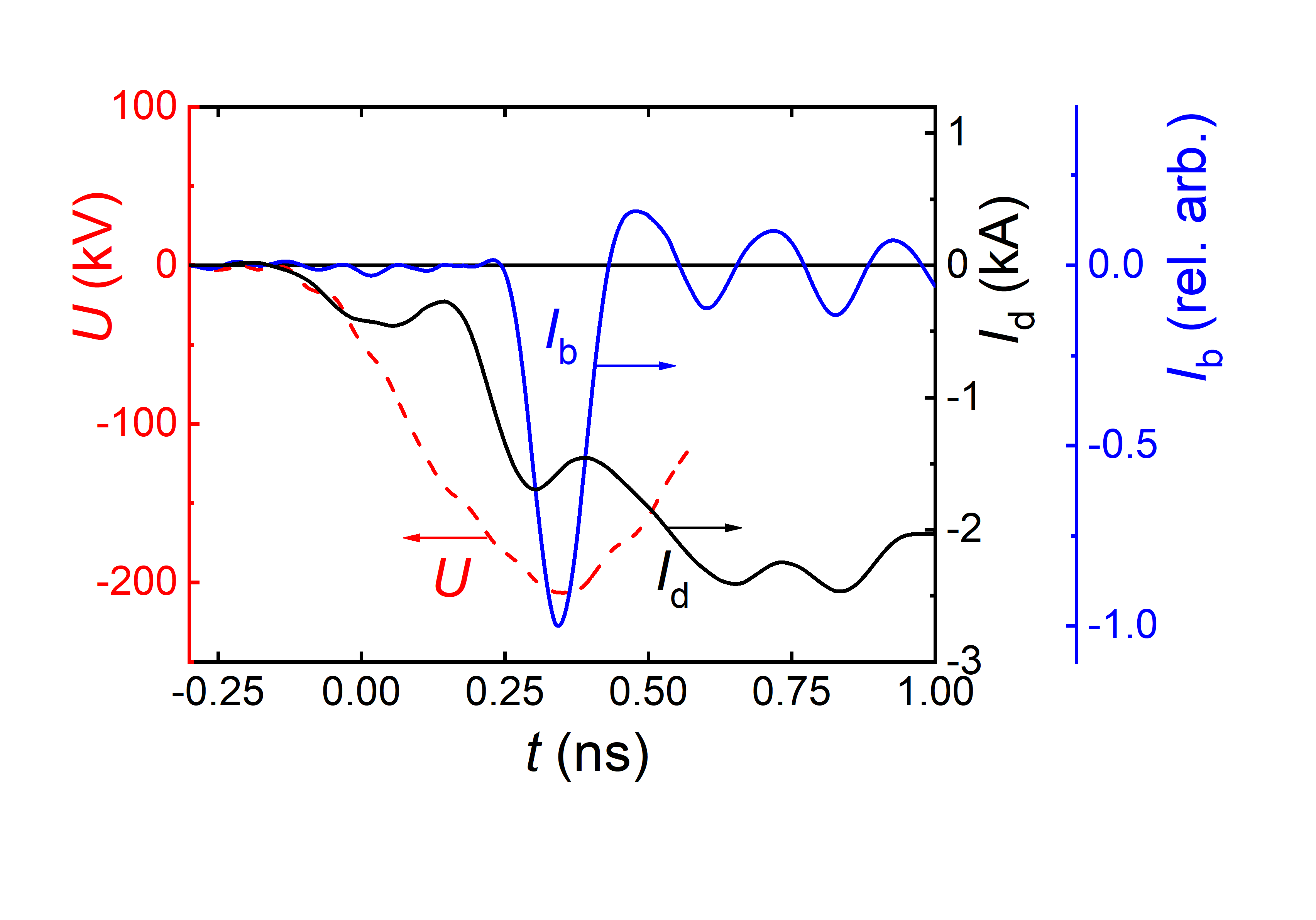}
\caption{\label{fig1} Waveforms of the voltage U, electron beam current at the SLEP-150M accelerator output $I_b$, and discharge current in the gas diode $I_d$. The cathode is a stainless steel ball, with a $6.5$~mm interelectrode gap. Setup~1a.}
\end{figure}

According to \cite{Tarasova, Babich, Baksht, Kozyrev_2015}, when high-voltage pulses with steep fronts are applied to a cathode with a certain radius of curvature in gas-filled diodes, the RE generation is observed. Oscillograms of the voltage pulses, beam current, and discharge current in air at atmospheric pressure, in Setup~1a from a diode with a spherical stainless-steel cathode, are shown in Fig.~\ref{fig1}. Note that the RE highest portion was recorded in experiments specifically with a cathode of this shape \cite{Baksht, Kozyrev_2015}. As can be seen, the generation of runaway electrons typically occurs at the leading edge of the voltage pulse and the discharge current. The RE beam was formed at a voltage amplitude of $200$~kV.

\begin{figure}[h]
\includegraphics[scale=0.37]{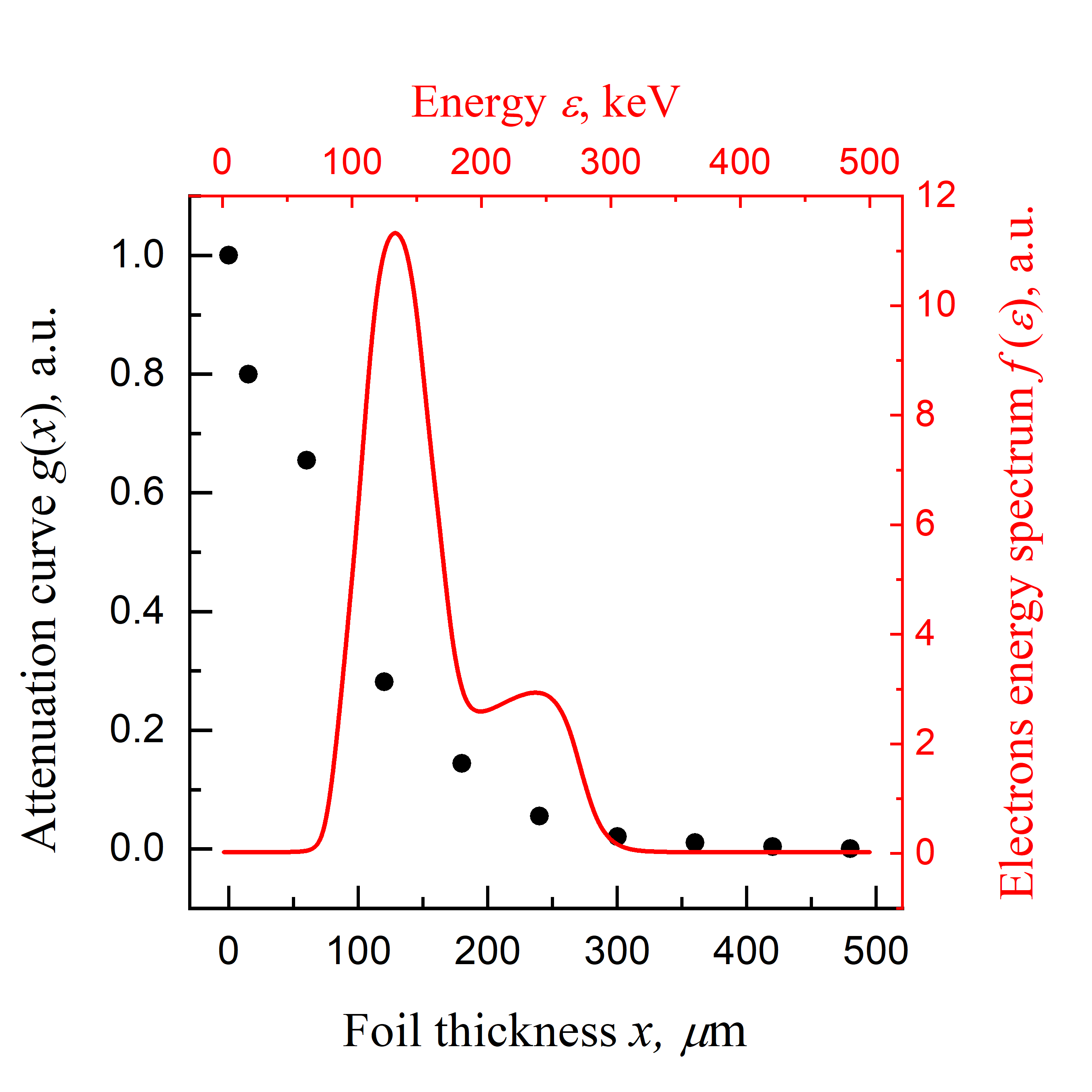}
\caption{\label{fig2} Electron beam attenuation coefficient (black dots) and electron spectrum at the collector (red curve) at atmospheric air pressure in a gas-filled diode. Setup~1a.}
\end{figure}

The beam current pulse duration at the receiving section of the $20$~mm-diameter collector was $100$~ps. The current pulse amplitude depends on the thickness and number of foils between the gas-filled diode and the collector. When calculating the RE spectrum from attenuation-curve data, a regularization technique was used \cite{Kozhevnikov_2026}. The beam current attenuation coefficient, together with the reconstructed RE spectrum, is shown in Fig.~\ref{fig2}.

Fig.~\ref{fig2} indicates that about $20$~\% of the beam electrons have energy greater than $eU$. An improved procedure for reconstructing the electron spectrum showed that the number of AEE in the beam spectrum is approximately twice that predicted by previous calculations using a simplified program; see \cite{Baksht, Kozyrev_2015}. Fig.~\ref{fig2} also shows that electrons with energy $>300\ keV$, if present, constitute a very small fraction ($<1$~\%).

\begin{figure}[h]
\includegraphics[scale=0.3]{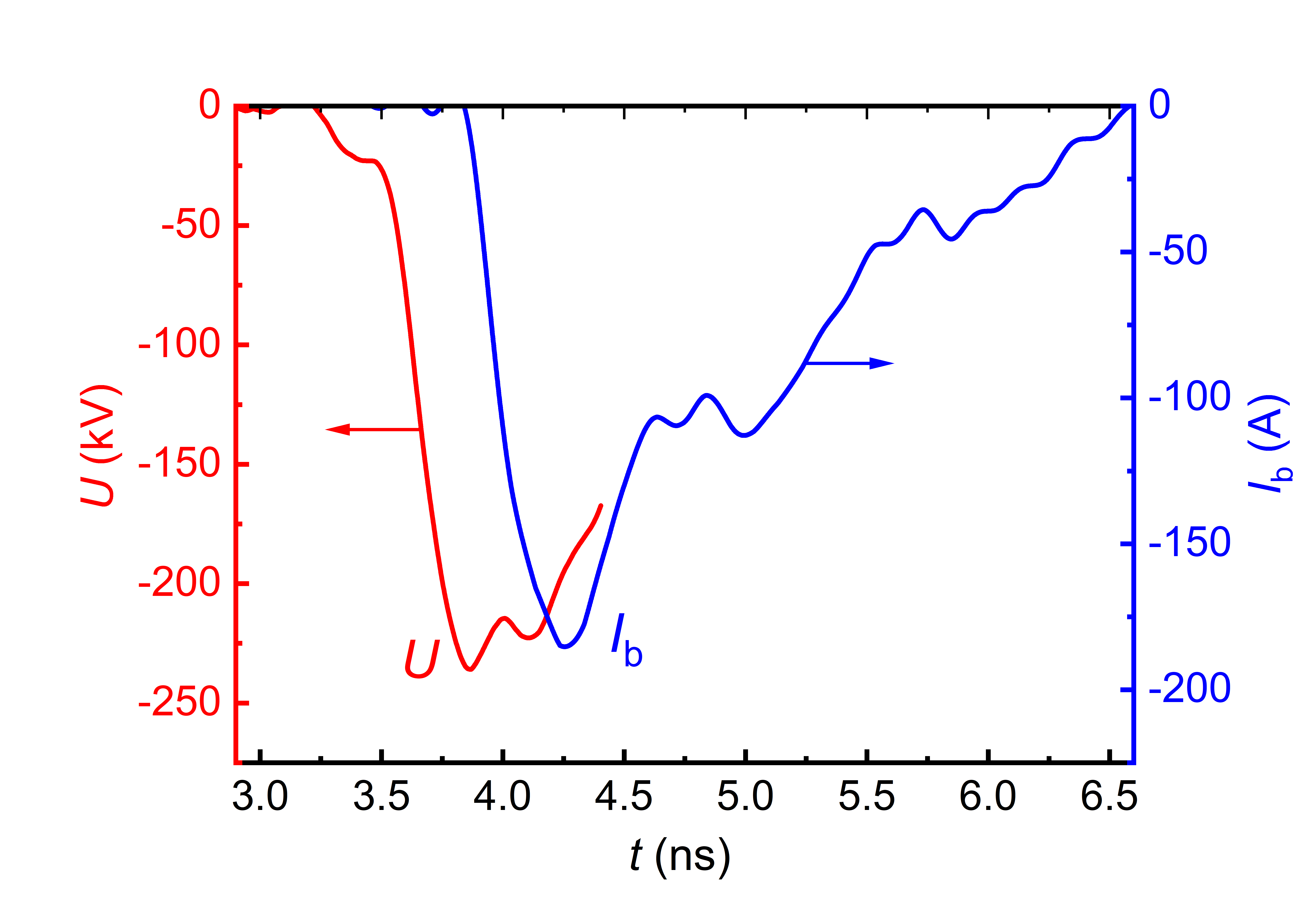}
\caption{\label{fig3} Waveforms of the voltage and beam current at the output of the SLEP-150M accelerator for a vacuum diode with a tubular cathode and $4$~mm interelectrode gap. Setup~1b.}
\end{figure}

For comparison, experiments were conducted on the same equipment for Setup~1b, using a vacuum diode with a tubular cathode and a $4$~mm interelectrode gap as the load. Oscillograms of the electron beam voltage and current pulses, obtained on the SLEP-150M accelerator with a vacuum diode, are shown in Fig.~\ref{fig3}. As can be seen, the voltage amplitude across the vacuum diode ($230$~kV) was comparable to that observed for a gas-filled diode. Meanwhile, the electron current in the vacuum diode (Fig.~\ref{fig3}) was an order of magnitude lower than the discharge current in the gas-filled diode (Fig.~\ref{fig1}).

\begin{figure}[h]
\includegraphics[scale=0.35]{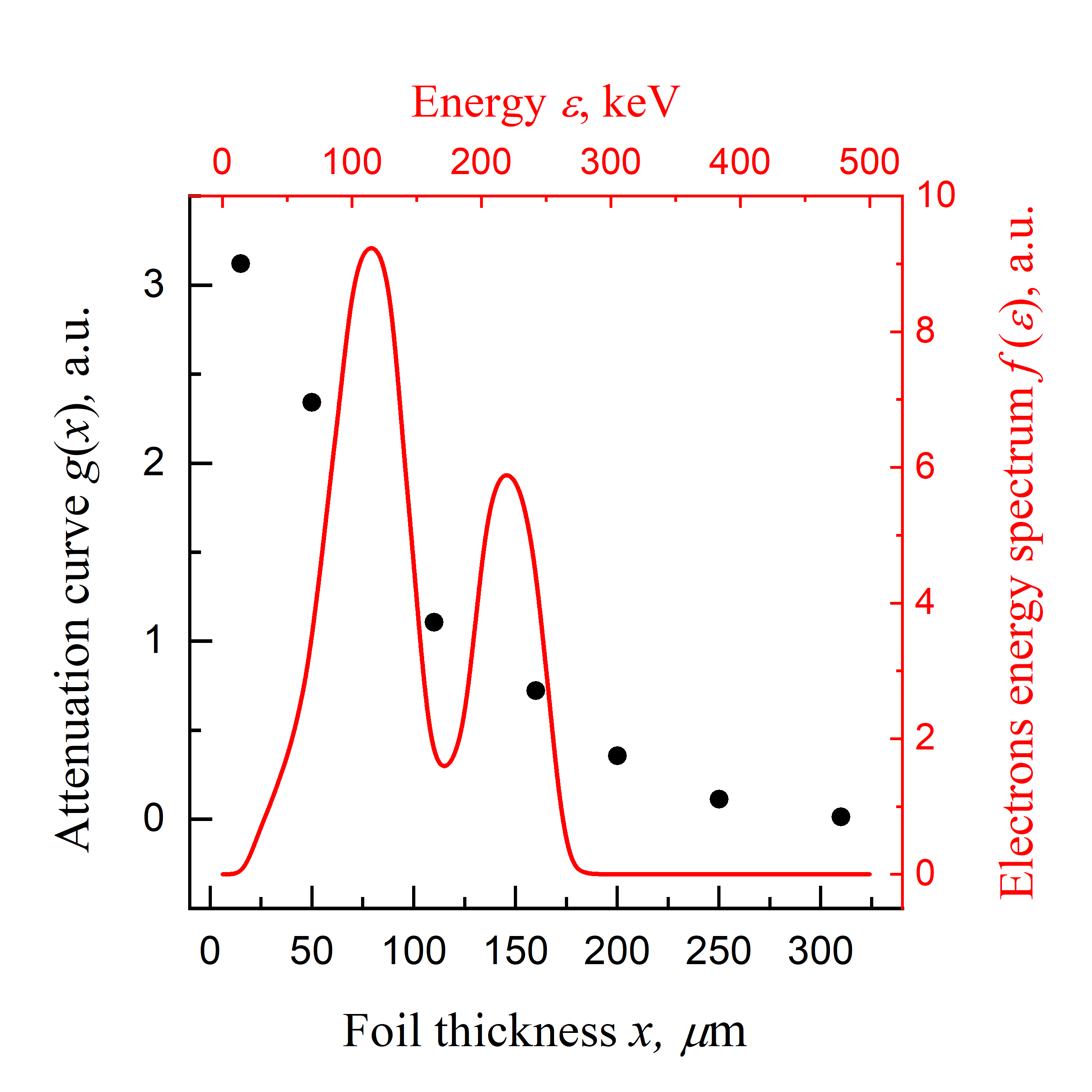}
\caption{\label{fig4} Electron beam attenuation curve (black dots) and electron spectrum at the output of the SLEP-150M accelerator with a vacuum diode (red curve), interelectrode gap of $4$~mm. Setup~1b.}
\end{figure}

The electron beam was generated at a diode voltage of $230$~keV; increasing the diode current decreased the gap voltage. The beam current pulse duration at the base was $3\ ns$. Under these conditions, the beam current attenuation coefficient and the calculated electron spectrum after foil filters are shown in Fig.~\ref{fig4}. The spectrum has two distinct maxima at $110$ and $230$~keV. It is evident that the vacuum diode also contains a group of AEE with energies higher than $eU\ =\ 230$~keV, whose relative portion is approximately $13$~\%.

The discontinuity in the voltage curves in Fig.~\ref{fig2} and Fig.~\ref{fig4} is due to the short transmission line length (the rapid arrival of the reflected pulse from the load at the capacitive divider), which prevented the recovery time from being extended further. However, in both cases, the voltage across the gap at the moment of AEE generation was determined. Thus, with a nanosecond voltage rise time in both vacuum- and gas-filled diodes, electrons with energies exceeding $eU$ are generated. Note that their percentage is significantly higher than the value estimated by the PEAB mechanism in \cite{Pasko_2024}.

To determine the time point at which AEE are generated in a vacuum diode, measurements of the collector current pulse profiles were conducted in Setup~2, which used a RADAN-220 generator with a voltage pulse duration of $2$~ns at FWHM. The RADAN-220 generator was chosen for its longer voltage pulse duration compared to the SLEP-150 generator. The load in Setup~2 was an IMA3-150E industrial vacuum diode with a voltage amplitude of $U\ =\ 200$~kV, and the electron beam was extracted into the atmosphere through a $300\ \mu$~m-thick beryllium anode foil. 

\begin{figure}[h]
\includegraphics[scale=0.3]{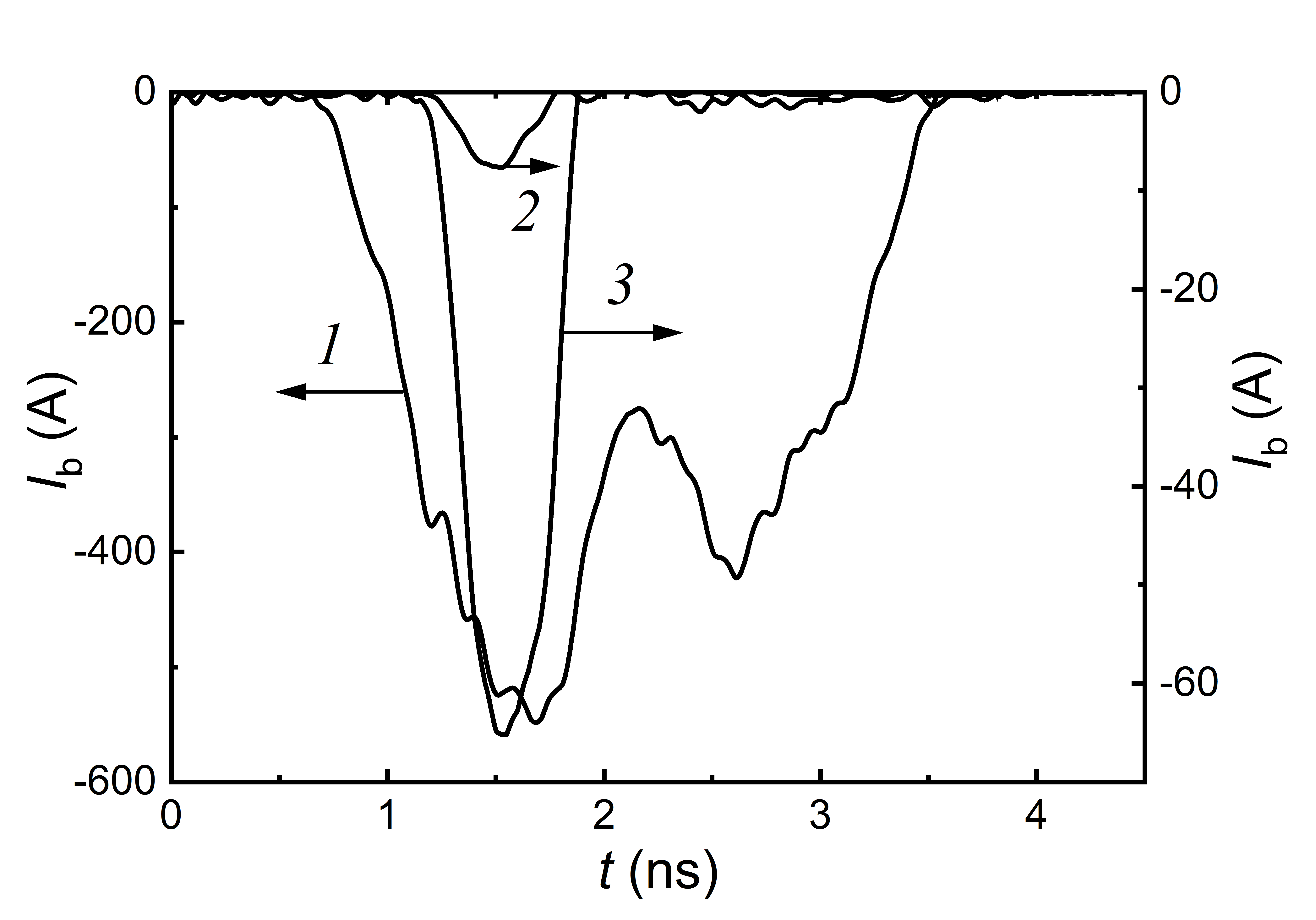}
\caption{\label{fig5} Waveforms of the electron beam behind the anode, synchronized in time, without a filter (1) and with additional aluminum filters of thicknesses $320\ \mu$m (2) and $250\ \mu$m (3). Setup~2.}
\end{figure}

At the start, the beam current was measured with a collector near the tube's output window. This current profile is shown in Fig.~\ref{fig5} as line 1; the beam current amplitude was $550$~A with a FWHM pulse duration of $1$~ns. Then, a $320\ \mu$~m thick aluminum foil was placed between the vacuum tube window and the collector; it allowed only electrons with kinetic energies $T\ >\ 260$~keV to pass through. The collector current pulse is shown in Fig.~\ref{fig5} as line 2; the current amplitude was $~\ 7$~A with a FWHM pulse duration of $~\ 250$~ ps. In an intermediate measurement variant, a $250\ \mu$~m thick aluminum foil was placed between the tube window and the collector; it allowed only electrons with energies $T\ >\ 240$~keV to pass through. The collector current pulse is shown in Fig.~\ref{fig5} as line 3; the current amplitude was $~\ 65$~A with a FWHM pulse duration of $~\ 450$~ps. It is now clear that electrons with energy $T\ >\ eU \approx\ 200$~keV are generated at the leading edge of the total beam current, as in the AEE in \cite{Baksht_2007}.

The experiments above demonstrated that AEE generation in both vacuum- and gas-filled diodes is attributed to the steep (subnanosecond) leading edge of the voltage pulse and possibly to a nonuniform electric field in the gap.

In \cite{Pasko_2024}, the diode configuration analyzed for PEAB was modeled as two parallel-plate electrodes. Our experiments most closely resembled this vacuum diode setup, specifically in Setup~3, which employed a plasma cathode. The diode's working gas pressure was approximately $10^{-2}$~Pa, ensuring that rare-particle collisions had minimal impact on electron energy within the accelerating gap. The cathode and anode were large, flat plates with a $12$~cm gap between them. A steady voltage of $150$~kV was applied across the diode. The total beam current reached $12$~A, with a pulse duration of $35\ \mu$s. In Setup~3, the perforated anode was fully covered with foil and measured $75 \times 15$~cm. The cathode grid, which extracted electrons from the plasma, had identical dimensions.

\begin{figure}[h]
\includegraphics[scale=0.35]{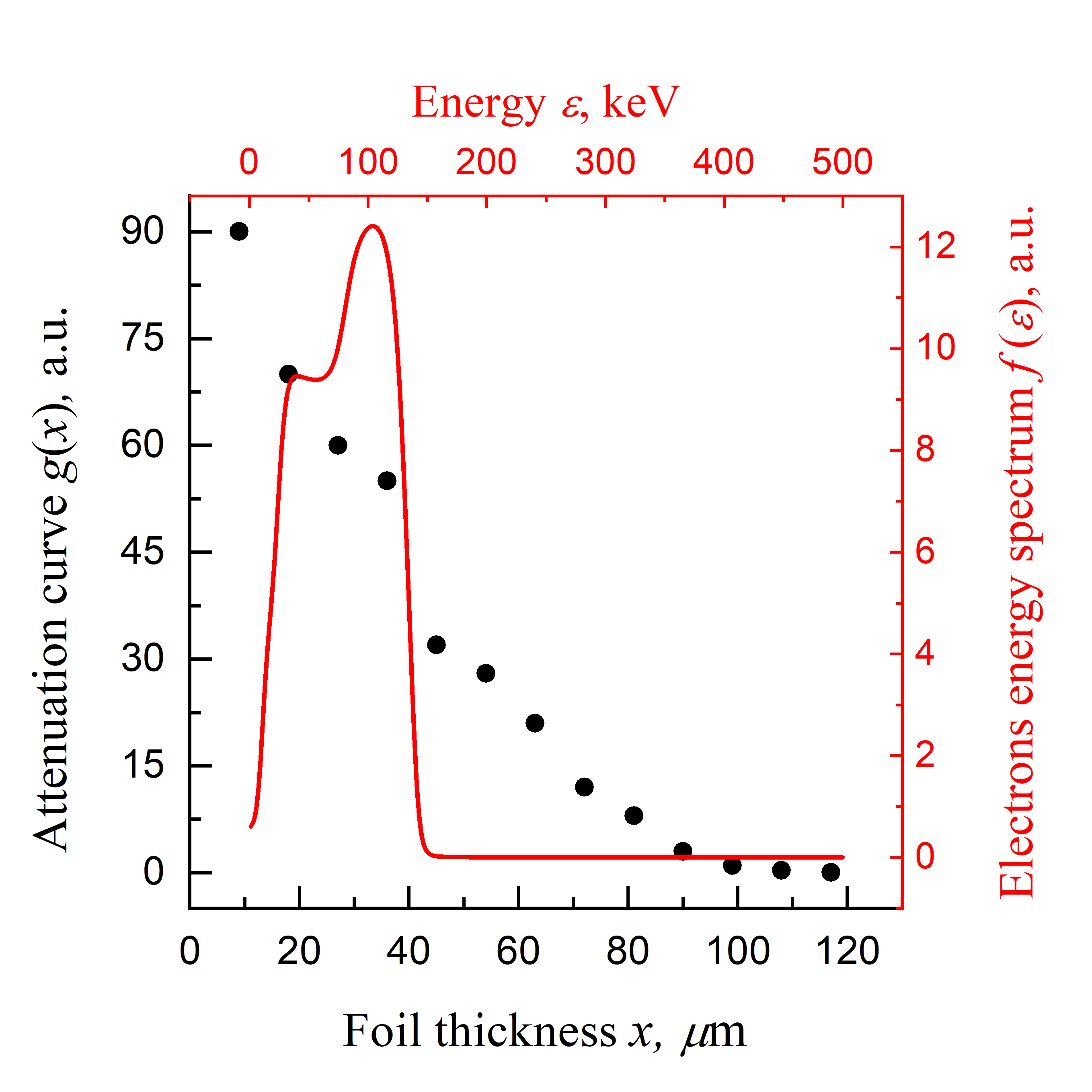}
\caption{\label{fig6} Electron beam attenuation curve (black dots) and electron spectrum (red curve) at the output of the accelerator of Setup~3 at a vacuum diode voltage of $150$~kV.}
\end{figure}

The long beam current duration and the large electrode area in experimental Setup~3 should facilitate the PEAB manifestation proposed in \cite{Pasko_2024} to explain the AEE in electrical discharges. The experimental data on the electron beam attenuation curve and the reconstructed electron spectrum are shown in Fig.~\ref{fig6}.

As shown in Fig.~\ref{fig6}, the electron spectrum drops to nearly zero at $150$~keV. Calculations showed that the AEE portion (with energies higher than $150$~keV) at Setup~3 was only $0.15$~\%. This value is significantly smaller than the error in reconstructing the electron spectrum, which did not exceed $1$~\% in the reconstruction method. In \cite{Pasko_2024}, calculations showed that with PEAB, the AEE should be approximately $1\ \%$ at an applied voltage of $100$~kV and approximately $7\ \%$ at $1$~MV. That is, at $150$~kV, this share should clearly be greater than $1\ \%$, but our measurements do not confirm this.

\section{Discussion and conclusions}

In \cite{Pasko_2024}, a chain of secondary processes explained the AEE generation of electrons: an electron emitted at the cathode - X-ray emission during the deceleration of a fast electron at the anode - emission of a fast electron at the cathode during the absorption of an X-ray quantum, and the cycle repeats. The secondary photoelectron, starting from the cathode with noticeable kinetic energy, gains the energy of the potential difference as it travels to the anode, arriving at the anode with an anomalous energy $> eU$. This mechanism is inevitably realized at a steady-state (or long-term) voltage on a vacuum diode; the only question is how to quantitatively estimate the number of AEE generated in this chain of processes. The problem is that the authors of \cite{Pasko_2024} compare results obtained for a vacuum diode with those of \cite{Tarasova, Tarasenko_2020}, in which AEE were measured in gas-filled diodes. In this paper, we demonstrate that the mechanism for the AEE generation in gas-filled diodes is completely different. This is convincingly demonstrated by our experiments and the fast-electron spectra derived from processed measurements. When the gap between the electrodes is filled with gas, the average electron can no longer acquire the energy of the potential difference it traverses, since kinetic energy losses from frequent inelastic collisions become decisive. At atmospheric pressure, a short electron beam is formed only in special cases described in \cite{Tarasova, Babich, Baksht, Kozyrev_2015}. The amount of AEE in the experiment (Fig.~\ref{fig2}) was less than $25\ \%$, which is approximately one order of magnitude higher than the estimates \cite{Pasko_2024} in the absence of collisions in the interelectrode gap.

We compare the AEE number estimates for a flat vacuum diode, based on the generation mechanism proposed in \cite{Pasko_2024}, with our experiments under a long voltage pulse applied to the vacuum diode (Fig.~\ref{fig6}). Here, the situation contrasts with that of the gas-filled diode: in the experiment, the AEE portion was significantly lower than the estimates by Pasko et al. Our data indicate an AEE below $0.2\ \%$, while our colleagues' estimates indicate a higher AEE of $0.8\ \%$. However, this difference is not as significant and can be easily explained by both the theoretical model's approximations and the uncertainty in the experimental measurements.

It has been shown that, in an atmospheric-pressure diode, the AEE portion can reach $25\ \%$. According to the "polarization" mechanism of electron acceleration at the streamer front \cite{Askaryan1, Askaryan2}, electrons acquire additional energy during their synchronous motion with the fast front of a dense plasma, which carries a region of locally high electric field strength. Our experiments on AEE generation are quantitatively consistent with the theoretical model of this mechanism presented in \cite{Kozyrev_2018}.

In vacuum diodes with explosive-emission cathodes, AEE are also observed at subnanosecond voltage rise times; "excessive acceleration" occurs due to the formation of a high-charge-density region of emitted electrons in the gap \cite{Boichenko, Kozhevnikov_2022}. It has been shown that with a nanosecond-duration voltage pulse, the AEE portion reaches $13\ \%$.

During long-duration voltage pulses (several $\mu$s) in a wide-aperture vacuum diode, which are most suitable for PEAB implementation \cite{Pasko_2023, Pasko_2024}, the AEE portion was found to be very small. This is also indirectly evidenced by the lack of data on AEE generation in studies \cite{Basov_1993, Sethian_1997, Bychkov_2000} focused on the development of wide-aperture electron accelerators. The measured energy spectra of electron beams show that, under the necessary conditions, AEE are generated in gas-filled and vacuum diodes, and their portions can exceed $10\ \%$. The necessary conditions are that voltage pulses have leading edges of $1$~ns or less. 

\begin{acknowledgments}
This research has been supported by the program of State assignment of the ISE SB RAS, project No.~FWRM-2026-0008, FWRM-2026-0009.
\end{acknowledgments}


\bibliography{apssamp}

\end{document}